\documentclass{article}
\usepackage{ltexpprt}

\usepackage{amsmath}
\usepackage{url}
\usepackage[pdfborder={0 0 0}]{hyperref}
\usepackage{cite}
\usepackage{graphicx}
\usepackage{booktabs}
\usepackage{subfig}

\graphicspath{{figures/}}

\newtheorem{observation}[lemma]{Observation}

\newcommand{\UnaryOperator}[2][]{%
	\ifx&#1&%
	\ensuremath{\mathop{}\mathopen{}#2\mathopen{}}%
	\else%
	\ensuremath{\mathop{}\mathopen{}#2\mathopen{}\left(#1\right)}%
	\fi%
}
\newcommand{\UnaryArray}[2][]{%
  \ifx&#1&%
  \ensuremath{\mathop{}\mathopen{}#2\mathopen{}}%
  \else%
  \ensuremath{\mathop{}\mathopen{}#2\mathopen{}\lbrack#1\rbrack}%
  \fi%
}
\newcommand{\Oh}[1]{\UnaryOperator[#1]{\mathcal{O}}}

\newcommand{\Text}[1]{\UnaryArray[#1]{\mathsf{T}}}
\newcommand{\Pattern}[1]{\UnaryArray[#1]{\mathsf{P}}}

\newcommand{\SA}[1]{\UnaryArray[#1]{\mathsf{SA}}}
\newcommand{\SAi}[2]{\UnaryArray[#1]{\mathsf{SA}_{#2}}}
\newcommand{\LCP}[1]{\UnaryArray[#1]{\mathsf{LCP}}}
\newcommand{\lcp}[2]{\ensuremath{\mathop{\mathsf{lcp}}\left( #1,#2\right)}}

\newcommand{\rank}[2]{\ensuremath{\mathop{\mathsf{rank}}_{#1}\!\left(#2\right)}}
\newcommand{\select}[2]{\ensuremath{\mathop{\mathsf{select}}_{#1}\!\left(#2\right)}}
\newcommand{\findclose}[1]{\UnaryOperator[#1]{\mathop{\mathsf{find\_close}}}}
\newcommand{\outdegree}[1]{\UnaryOperator[#1]{\delta^+\!}}
\newcommand{\pathlabel}[1]{\UnaryOperator[#1]{\mathsf{pathlabel}}}
\newcommand{\lab}[1]{\UnaryOperator[#1]{\mathsf{label}}}
\newcommand{\abs}[1]{\ensuremath{\left\vert#1\right\vert}}
\newcommand{\strdepth}[1]{\UnaryOperator[#1]{\mathsf{sd}}}

\newcommand{\CN}{PE}
\newcommand{\GT}[1]{\UnaryOperator[#1]{\mathsf{GT}}}
\newcommand{\PT}[1]{\ensuremath{\mathsf{PT}}}
\newcommand{\DPT}[1]{\ensuremath{\mathsf{DPT}}}
\newcommand{\DSA}[1]{\ensuremath{\mathsf{DSA}}}
\newcommand{\PatternSize}{\ensuremath{\abs{\mathsf{P}}}}
\newcommand{\TimeTrie}[1]{\UnaryOperator[#1]{t_{\text{trie}}}}
\newcommand*\samethanks[1][\value{footnote}]{\footnotemark[#1]}

\begin{document}

\title{\Large Engineering a Distributed Full-Text Index\thanks{This work was supported by the German Research Foundation (DFG), priority programme ``Algorithms for Big Data'' (SPP 1736).}}

\author{Johannes Fischer\thanks{Technische Universit{\"a}t Dortmund, Department of Computer Science, \href{mailto:johannes.fischer@cs.tu-dortmund.de}{johannes.fischer@cs.tu-dortmund.de}, \href{mailto:florian.kurpicz@tu-dortmund.de}{florian.kurpicz@tu-dortmund.de}} \and Florian Kurpicz\samethanks \and Peter Sanders\thanks{Karlsruhe Institute of Technology, Institute for Theoretical Informatics, \href{mailto:sanders@kit.edu}{sanders@kit.edu}}}

\date{}
\maketitle

%\pagenumbering{arabic}
%\setcounter{page}{1}%Leave this line commented out.

\begin{abstract}
  \small\baselineskip=9pt
  We present a distributed full-text index for big data applications in a distributed environment.
  Our index can answer different types of pattern matching queries (existential, counting and enumeration).
  We perform experiments on inputs up to 100 GiB using up to 512 processors, and compare our index with the distributed suffix array by Arroyuelo et al.~[Parall.~Comput.\ 40(9): 471--495,~2014].
  The result is that our index answers counting queries up to $5.5$ times faster than the distributed suffix array,
  while using about the same space.
  We also provide a succinct variant of our index that 
  uses only one third of the memory compared with our non-succinct variant, at the expense of only 20\%\ slower query times.
\end{abstract}

\section{Introduction} % (fold)
\label{sec:intoduction}

Index data structures are one of the most powerful tools for coping with large data sets.
Among the index data structures for texts are suffix arrays, suffix trees and related structures that allow full-text search of patterns in time \emph{independent} of the size of the text corpus.
Consequently, there are thousands of papers on such suffix data structures.
However, when you look at truly large data sets that do not fit on a single machine, there is very little work yet.
We found that very surprising since the biggest inputs are those where the index has the biggest impact.
Refer to \S~\ref{sub:related_work} for a more detailed discussion of related work.
You can easily adopt the approach from commercial search engines (which use inverted index data structures) to distribute your corpus over the machines and then use a local index on each machine.
However, then the amount of work and energy you invest in a query grows proportional to the number of machines and thus, \emph{linearly} with the corpus size -- you have (asymptotically) thrown away the huge advantage of a powerful index data structure.

Our contribution is the development of a truly distributed full-text index data structure that supports typical queries by exchanging only a constant number of messages whose length is proportional to the length of the search pattern.

\subsection{Overview.}
Our construction algorithm starts from a distributed input text $\Text{}$ and a distributed suffix array (\SA{}) together with information on the longest common prefix of subsequent entries in the \SA{}, the so called \LCP{}-array.
In \S~\ref{sec:preliminaries} we formally introduce the \SA{}, \LCP{}-array, and other prerequisites.
By scanning the \SA{} and \LCP{}-array, we then construct a two-level trie data structure.
Using succinct data structures we can reduce the size of the trie to 15 bits per character of the text.
In our scenario, a \emph{processing elements} (\CN{}) denotes a processor on a compute node in the cluster.
A small top-level trie \GT{} (see \S~\ref{ssub:global_trie} for more details) is sufficient to decide which \CN{}s are involved in answering a query.
\GT{} is replicated over all \CN{}s so that queries can arrive anywhere and get forwarded to those \CN{}s that hold the relevant part of the \SA{}.
The search in the local part of a \SA{} is facilitated by a succinctly represented Patricia trie on that part.
After this local search, a single remote access to the text suffices to locate the pattern in the \SA{}.
Refer to \S~\ref{ssub:local_tries} for a more detailed explanation of our data structure.

\subsection{Related Work.} % (fold)
\label{sub:related_work}
Multi-level full-text indices have been considered for external memory. The \emph{String B-Tree}~\cite{Ferragina1999} utilizes Patricia tries at each level to reduce the I/O volume.
There exists also theoretical work by Ferragina and Luccio~\cite{Ferragina1999a} that discusses a distributed Patricia trie.
Their approach is only good when answering \emph{long} queries, for example, existential queries ``does the pattern occur in the text?'' of length $m\geq c$, where $c$ is the number of \CN{}s.
Those queries can be answered optimally with respect to computation and communication.
For real world applications this does not provide a satisfactory solution, as the average length of natural queries is shorter (18 AOL~\cite{Pass2006} and 21 TREC~\cite{MillionQueryTrack} on average) than the number of \CN{}s we want to utilize.
% awk ' { totlen+=length($0); } END {
% printf("average: %d\n", totlen/NR); } ' FILE_NAME

A complementary (theoretical) approach is described by M{\"{a}}kinen et al.~\cite{Makinen2004} and is good for short patterns.
Using backwards search, a query can be answered using $m$ communication steps.
The problem is that at most $\sigma$ \CN{}s can be used and, in the
worst case, some \CN{}s might need space $\Omega(n)$.

Arroyuelo et al.~\cite{Arroyuelo2014} consider a large variety of distributed suffix array data structures with various tradeoffs between replication, number of remote data accesses, and load balance.
However, they always use explicit binary search in the suffix array, leading to logarithmically higher costs than our approach.
Our index can be viewed as an improvement of the \emph{global} approach presented in~\cite{Arroyuelo2014}, where we add two levels of tries that allow us to answer each query exchanging only a constant number of messages.
We show that our index scales better especially for larger text sizes and a larger number of \CN{}s.

Russo et al.~\cite{Russo2010} (theoretically) describe distributed compressed indices. Their approach partitions the text between the \CN{}s and works with local indices.
The consequence is that queries have to be processed on every \CN{}, contrary to our goal to have total work independent of $c$.

It is known that the suffix array can be computed efficiently in parallel~\cite{Karkkainen2006,Kulla2007}.
Flick et al.~\cite{Flick2015} give the only distributed algorithm for computing both the \SA{} and \LCP{}-array; their approach is within a factor of $\Oh{\log n}$ from the optimal.
We use the \SA{} and \LCP{}-array as the starting point for our index construction and we do not study their construction in this paper.
Note that none of these papers~\cite{Karkkainen2006,Kulla2007,Flick2015} discuss how to actually use the resulting data structure for a distributed query.

There are several results that assume that the input text is replicated over all \CN{}s (e.g.~\cite{Clifford2005}).
This makes index construction and search much easier but severely limits scalability, so we do not consider them further.
% subsection related_work (end)
% section introduction (end)

\begin{figure*}
  \centering
  \subfloat[\label{sfig:succinct_representations}]{%
    \includegraphics{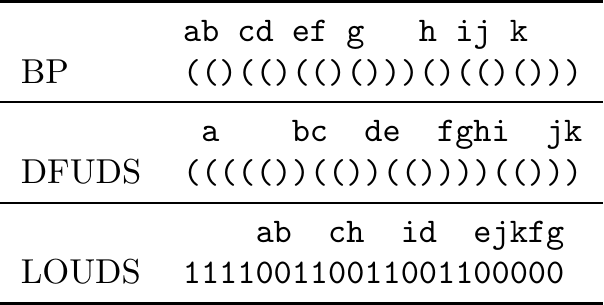}
  }~~~
  \subfloat[\label{sfig:tree_structure}]{%
    \includegraphics[height=10\baselineskip]{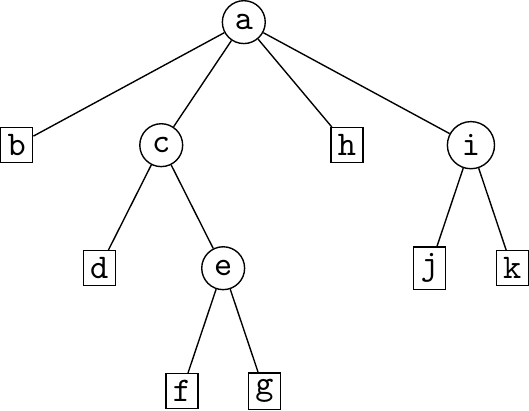}
  }
  \caption{\protect\subref{sfig:succinct_representations}~Succinct representations of the tree shown in~\protect\subref{sfig:tree_structure}.
  \label{fig:succinct_examples}}
\end{figure*}

\section{Preliminaries} % (fold)
\label{sec:preliminaries}
\subsection{Suffix Arrays.}
Let $\Text{}=\Text{1}\Text{2}\dots\Text{n}$ be a text of length $n$ over an alphabet $\Sigma$ of size $\abs{\Sigma}=\sigma$.
$\Text{i..j}$ denotes the \emph{substring} $\Text{i}\dots\Text{j}$ for all $1\leq i\leq j\leq n$.
Also, $P_i=\Text{1..i}$ is $i$-th \emph{prefix} of \Text{} and $S_i=\Text{i..n}$ is the $i$-th \emph{suffix} of \Text{} for all $1\leq i\leq n$.
The longest common prefix (\textsf{lcp}) denotes the maximum size of a common prefix of two suffixes of the text, i.e., $\lcp{i}{j}=\max\lbrace k\geq 0:\Text{i..i+k-1}=\Text{j..j+k-1}\rbrace$ for all $1\leq i,j\leq n$.

The \emph{suffix array} \cite{Manber1993} \SA{} of a text \Text{} with $\abs{\Text{}} = n$ is a permutation of $\lbrack 1,n\rbrack$ such that it enumerates the starting positions of the suffixes of \Text{} in lexicographical order.
Hence, $S_{\SA{i}}<_{\text{lex}}S_{\SA{j}}$ for all $1\leq i< j\leq n$.
The \emph{longest common prefix array} (\LCP{}-array) enhances the \SA{} and contains the size of the longest common prefix of two lexicographically consecutive suffixes, i.e., $\LCP{1}=0$ and $\LCP{i}=\lcp{\SA{i-1}}{\SA{i}}$ for all $2\leq i\leq n$.

The \SA{} and \LCP{}-array are the foundation of the indices considered in this paper, i.e., the index we present and the index we compare our results with.
Not only can they be simulated using these two arrays, but they can also be constructed using only the \SA{}, \LCP{}-array and \Text{}.
As already mentioned in the introduction, there exists a distributed algorithm for constructing the \SA{} and \LCP{}-array~\cite{Flick2015}, which is a distributed variant of Larsson and Sadakane's algorithm~\cite{Larsson2007}.
Using $c$ \CN{}s it can compute the \SA{} in $\Oh{t_\text{sort}\left(n,c\right)\lg n}$ time, where $t_\text{sort}\left(n,c\right)$ denotes the time required to sort $n$ elements that are uniformly distributed over $c$ \CN{}s.
The \LCP{}-array construction requires $\Oh{\left(\frac{n}{c}+c\right)\lg n}$ additional time.
Another distributed \SA{} construction algorithm is the pDC3~\cite{Kulla2007}, which is a distributed variant of the DC3 algorithm~\cite{Karkkainen2006}, can compute the \SA{} in time $\Oh{\frac{n\lg n}{c}+\lg^2 c}$, i.e., a log-factor better than~\cite{Flick2015}.
For our experiments we use the implementation of the latter, as it is the only available distributed \SA{} and \LCP{}-array construction algorithm.

\subsection{Tries.}
Given a labeled tree $G=\langle V,E\rangle$ with root $r\in V$, we denote the label of a node or an edge by \lab{\cdot} and the concatenation of all edge labels on the path from the root to a node $v$ by \pathlabel{v}.
The \emph{out-degree} of a node $v$ is denoted by \outdegree{v}.
The \emph{leaf rank} of a leaf $\ell\in V$ is the number of leaves visited before $\ell$ in a preorder traversal of the tree.

Let $R=\lbrace R_1,R_2,\dots,R_k\rbrace$ be a set of strings over the alphabet $\Sigma$ such that all strings are distinct and no string is the prefix of another string in $R$.
The \emph{trie} of $R$ is an ordered tree with root $r$, where the edge labels are characters and the leaves represent string numbers from $[1,k]$ such that:
\begin{enumerate}
\item for each node $v\in V$, the labels of the outgoing edges
  $\lab{\left(v,\cdot\right)}\in\Sigma$ are distinct,
\item for each string $R_i\in R$, there is a leaf $\ell\in V$ with
  $R_i=\pathlabel{\ell}$ and $\lab{\ell}=i$ and
\item for each leaf $\ell\in V$ there is a string $R_i\in R$ such that
  $R_i=\pathlabel{\ell}$ and $\lab{\ell}=i$.
\end{enumerate}

The \emph{compressed trie} is a trie where each path $e_1,e_2,\dots,e_\ell$ with $\ell>1$ consisting only of nodes with out-degree $1$ is replaced by a single edge $e$ such that $\lab{e}=\lab{e_1}\lab{e_2}\dots\lab{e_\ell}$.
Still, all outgoing edges of a node $v$ start with a different character.
The \emph{string depth} of a node $v$ is $\strdepth{v}=\abs{\pathlabel{v}}$, i.e., the length of the longest common prefix of all strings represented leaves below $v$.
To find all occurrences of a pattern \Pattern{} in a compressed trie, we start at the root $r$ and follow the edge $e$ such that $\lab{e}=\Pattern{1..\abs{\lab{e}}}$.
At each node $v$, the length of the pattern matched up to this point equals \strdepth{v}.
We then follow the edge $e$ with $\lab{e}=\Pattern{\strdepth{v}+1..\abs{\lab{e}}+\strdepth{v}}$.
This process is repeated until we have matched the whole pattern at the edge $\left(\cdot, v\right)$.
Then, all leaves that are successors of $v$ correspond to strings in $R$ that are prefixed by \Pattern{}.
If at any point, there is no edge to follow, the pattern~\Pattern{} does not occur in the trie.

The \emph{Patricia trie} (or \emph{blind trie})~\cite{Morrison1968} of a text \Text{} is a compressed trie for all suffixes of \Text{}, where each node $v$ just stores the first character and the string depth $\strdepth{v}$.
Due to this limitation, finding all occurrences of a pattern requires two steps -- a \emph{blind search} followed by a comparison to a substring of \Text{} (which has been determined by the blind search):

For the blind search, we start at the root and follow the edge matching the pattern at the position corresponding to the string depth, i.e., at a node $v$ we follow the edge $e$ with label $\lab{e}=\Pattern{\strdepth{v}}$.
We repeat this until we have reached a node $v$ such that $\strdepth{v}\geq\abs{\Pattern{}}$ or there is no feasible edge to follow.
In the first case, we retrieve a prefix of length $\abs{\Pattern{}}$ of a suffix corresponding to any leaf $w$ that is a successor of $v$ and compare that prefix with our pattern \Pattern{}.
In the second case (there is no edge to follow) \Pattern{} does not occur in \Text{}.

Next, we compare \Pattern{} and $\Text{i..i+\abs{\Pattern{}} - 1}$ where $i$ is the label of the leaf $w$ (that has been identified during the blind search).
If the strings are equal, then all leaves that are below $v$ correspond to an occurrence of \Pattern{} in \Text{}.
Otherwise, \Pattern{} does not occur in \Text{}.

The Patricia trie can be constructed from the \SA{}, \LCP{}-array and \Text{} in linear time, i.e., scanning the \SA{} and \LCP{}-array once and considering each entry at most twice.
\Text{} is required for the edge labels and each position is accessed at most once.
In \S~\ref{sub:construction} we give a detailed description of the construction algorithm and in \S~\ref{sub:experiments_construction} we compare the construction time for the tries needed by our index with the time required for the construction of the \SA{} and \LCP{}-array.

\subsection{Succinct Data Structures.}
We can represent a tree containing $\ell$ nodes using a bit vector $B\in\lbrace 0,1\rbrace^{2\ell}$.
The bits represent parentheses; a $1$ represents an open parenthesis ``$($'' and a $0$ represents a closing parenthesis ``$)$''.
To navigate in the tree, we require additional operations on the bit vector:
$\rank{0}{i}$ asks for the number of $0$'s in $B$ up to position $i-1$, $\select{0}{i}$ returns the position of the $i$-th $0$ in $B$, and $\findclose{i}$ gives the position of the matching closing parenthesis for an open parenthesis at position $i$.
The operations $\rank{1}{\cdot}$ and $\select{1}{\cdot}$ work analogously for $1$'s in the bit vector.
All these operations can be answered in constant time~\cite{Jacobson1989,Clark1996}.
The \emph{level ordered unary degree sequence} (LOUDS)~\cite{Jacobson1989} represents a tree level-wise, i.e., starting at the root, we visit all nodes $v$ of a level from left to right and add $\outdegree{v}$ $1$'s followed by a $0$ to the bit vector.
The position of the $i$-th child of the node at position $x$ is identified by $\select{0}{\rank{1}{x} + i - 1} + 1$ in constant time.
The \emph{depth first unary degree sequence} (DFUDS)~\cite{Benoit2005} is obtained by traversing the tree in preorder and (like in LOUDS) append $\outdegree{v}$ $1$'s followed by a $0$ whenever we visit a node $v$ for the first time.
To make the sequence balanced, we prepend a $1$.
The position of the $i$-th child of the node at position $x$ is identified by $\findclose{\select{0}{\rank{1}{x} + 1} + 1} + 1$ in constant time.
Last, the \emph{balanced parenthesis} (BP) representation~\cite{Munro2001} is also constructed by traversing the tree in preorder:
We add a $1$ to the bit vector whenever we visit a node for the first time and we add a $0$ to the bit vector whenever we visit a node for the last time.
In theory, BP also allows an access of the $i$-th child in constant time~\cite{Sadakane2010}.
However, in the implementation used by us (see \S~\ref{sec:experiments} for more details), BP does not support a direct access to the $i$-th child.
Instead, one has to access the first child of the node at position $x$ (position $x+1$) and then go to the next child ($\findclose{x} + 1$) until the $i$-th child is reached in \Oh{i} time.
In our implementation, the first two representations allow an access of the $i$-th child in in constant time, whereas it takes \Oh{i} time in BP.
See Figure~\ref{fig:succinct_examples} for an example of these succinct tree representations.

\subsection{Model of Computation.} % (fold)
\label{sub:theoretical_model}
In the \emph{bulk-synchronous parallel} (BSP) model~\cite{Valiant1990}, each computation is a sequence of \emph{supersteps}.
Each superstep is split into three parts.
First, the \CN{}s can perform any number of operations based on local data.
This is followed by a communication phase, where the \CN{}s can send data to other \CN{}s.
After the communication, all \CN{}s are synchronized, i.e., all \CN{}s wait until the last \CN{} has finished all operations on local data and communication.
There is no synchronization between the first and second part, \CN{}s can start communicating as soon as they have finished working on the local data (the results of the communication are not available during the superstep).
Then at the beginning of the next superstep, each \CN{} can use the data retrieved during the last superstep.
The total running time of a BSP program is the cost of all its supersteps, where the cost of one superstep is $w+hG+L$. Here,
\begin{itemize}
\item $w$ is the maximum time used for computation by each \CN{} (excluding communication),
\item $h$ is the maximum of machine words communicated by each \CN{}, with $G$ being the running time required for the communication of one machine word, and
\item $L$ is the cost of the barrier synchronization.
\end{itemize}
BSP has been considered as the model for other distributed indices \cite{Arroyuelo2014,Ferragina1999a}.
Also, we will see that the BSP model suits our setting well, as we have well defined phases where communication is required during construction and query processing -- see \S~\ref{sec:distributed_patricia_trie}.
Whenever we \emph{retrieve} data we refer to \emph{direct remote memory access} (DRMA) that is supported by the BSP model~\cite{Skillicorn1997}.
% subsection theoretical_model (end)

\label{ssub:pattern_matching_queries}
\begin{figure*}
  \centering
  \includegraphics[scale=1.125]{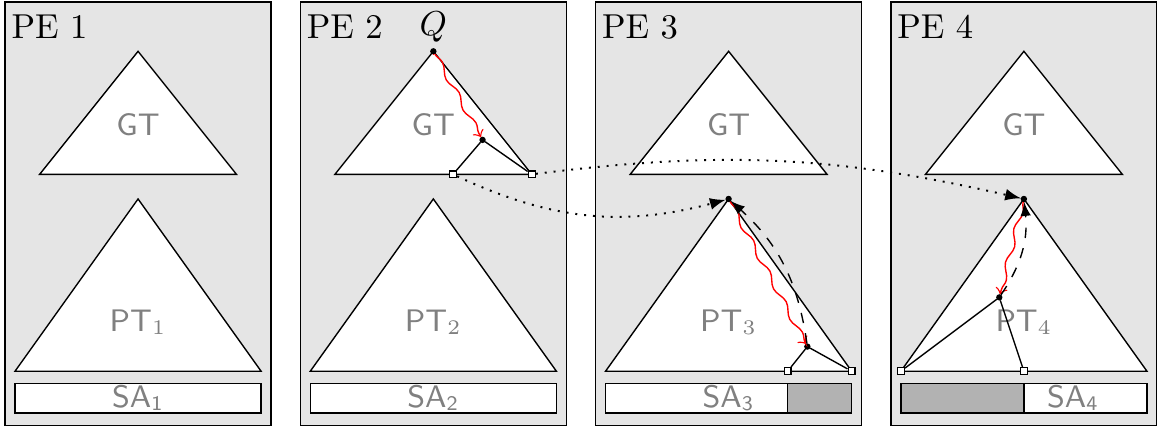}
  \caption{The first level of the \DPT{} is a trie (\GT{}) over the smallest and largest suffix stored by the local suffix array at each \CN{}.
  It is the same at each \CN{}.
  The lower level at \CN{} $i$ is the local Patricia trie $\PT{}_i$ over $\SA{}_i$.
  A query $Q$ is answered in four supersteps.
  First, the \CN{}s that are responsible for the query are determined (using the trie \GT{} at the first level).
  Then the query is sent to those \CN{}s.
  In the second superstep, a blind search in the Patricia trie $\PT{}_i$ (second level of \CN{} $i$) is executed.
  The substring corresponding to the result of the blind search is retrieved and in the third superstep the query is answered using that substring.
  % In the last superstep, the result is sent to the \CN{} that got the query initially.
  \label{fig:index_overview}}
\end{figure*}

\section{Distributed Patricia Trie} % (fold)
\label{sec:distributed_patricia_trie}
Let $\Text{}=\Text{1}\Text{2}\dots\Text{n}$ be a text of length $n$.
The \CN{}s are numbered from $1$ to $c$.
We assume that $n$ is divisible by $c$.
Then we distribute the \SA{} and \LCP{}-array in a consecutive fashion, such that the $i$-th \CN{} holds $\SA{}_i\ =\SA{1+\left(i-1\right)\frac{n}{c}..i\frac{n}{c}}$ and $\LCP{}_i\ =\LCP{1+\left(i-1\right)\frac{n}{c}..i\frac{n}{c}}$.
In addition, each \CN{} holds a part of the text as described in the next section.

Our proposed data structure, the \emph{distributed Patricia trie} (\DPT{}), is a two level index consisting of an index \GT{} for query distribution (first level) and several indices $\PT{}_i$ that can find all occurrences of a pattern \Pattern{} that starts at text positions held by the local $\SA{}_i$ on \CN{} $i$ (second level).
In this case we say that the \CN{} $i$ is \emph{responsible} for \Pattern{}.
The index \GT{} is replicated at every \CN{}.
This allows queries to arrive at arbitrary \CN{}s and then to be sent to the responsible \CN{}s in the next step.
There, the query is processed utilizing $\PT{}_i$.
This index is unique for each \CN{} -- see also Figure~\ref{fig:index_overview}.

\subsection{Construction.} % (fold)
\label{sub:construction}
In this section we show how to construct the \DPT{} in linear time.
We start with the construction of the local \PT{}s as we use the information about their smallest and greatest element for the construction of \GT{}.

\paragraph{Local Tries.}
\label{ssub:local_tries}
The construction is the same at each \CN{}.
Our construction algorithm is the extension of an algorithm to compute the \emph{suffix tree}, i.e., a compressed trie of all suffixes of a text \Text{}.
We modified the suffix insertion algorithm presented in~\cite[p.~143]{Makinen2015} such that the Patricia trie can be constructed by scanning the \SA{} and \LCP{}-array from left to right.
The pathlabel of the rightmost path in a (Patricia) trie is the lexicographically largest pathlabel in the trie.
Since all suffixes in \SA{} are in lexicographical order, each suffix that is added to the Patricia trie is lexicographical greater that all previously inserted suffixes and will form the new rightmost path.
Therefore, at each point of time during the construction, only nodes on the rightmost path can be changed.
All other nodes are considered as \emph{final}.
The inner nodes on the rightmost path, i.e., the nodes that can still be changed, are kept on a stack.
Since we compute a Patricia trie, each node $v$ knows its string depth \strdepth{v}.

Initially, we have a stack containing a node with string depth 0 and no children.
We start by adding the first inner node $v$, with $\strdepth{v}=\LCP{2}$, two children (the left child represents $\SA{1}$ and the right child represents $\SA{2}$) with edge labels $\Text{\SA{1} + \LCP{2}}$ and $\Text{\SA{2} + \LCP{2}}$, resp.
If $\strdepth{v}=0$, the node replaces the initial one that has been on the stack.
Otherwise, $v$ will be a child of the initially created node.
We now continue to scan the \SA{} and \LCP{}-array from left to right.
Whenever we read a new position $i$ in the \LCP{}-array, we remove nodes from the stack until the node $v$ on top of the stack has $\strdepth{v}\leq\LCP{i}$.
If $\strdepth{v}<\LCP{i}$, we create a new inner node $w$ with $\strdepth{w}=\LCP{i}$, i.e., we \emph{branch below} node $v$.
The left child of $w$ (edge label $\Text{\SA{i-1} + \LCP{i}}$) is the former rightmost child of $v$, and the right child of $w$ is a new leaf referring to $\SA{i}$ and has edge label $\Text{\SA{i} + \LCP{i}}$.
Next, $w$ becomes the new rightmost child of $v$ and is put on the stack.
If $\strdepth{v}=\LCP{i}$, $v$ just gets a new rightmost child (edge label $\Text{\SA{i} + \LCP{i}}$) referring to $\SA{i}$, i.e., $v$ gets a new \emph{leaf}.
Following these operations, we can compute each local \PT{} in \Oh{\frac{n}{c}} time.

With respect to practical application, we also want to construct succinct representations of the tries.
It is possible to compute a succinct representation using its pointer based representation.
Using the approach described above, we can also compute a succinct trie representation directly, i.e., reducing the required memory peak for the construction.
We compute the DFUDS representation of a trie by storing all final nodes and their subtrees in DFUDS representation.
Whenever we remove a node from the stack, we add it and its subtree at the end of the already computed DFUDS representation of the previously removed final nodes.
This is possible because we construct the trie in the same order as a depth first search traversal visits all nodes (which is the order in which the nodes are represented in DFUDS).

Up to now, we have simply named the characters that correspond to the edge labels.
Since all local \PT{}s are on different \CN{}s, we cannot assure that the text position required for an edge label is locally available.
We have to retrieve all edge labels during one communication phase.
The number of characters stored at each \CN{} is $\Theta(\frac{n}{c})$.
During the construction of the local Patricia trie $\PT{}_i$, we scan the arrays $\SA{}_i$ and $\LCP{}_i$ to determine the first mismatching text positions of two lexicographically consecutive suffixes.
These characters will then be used for the edge labels later on.
If we create a new leaf, we only require \emph{one} edge label.

For a simpler and more realistic analysis of the costs for constructing our local indexes, we assume that all mismatching characters are stored at the same \CN{} where the corresponding suffix starts, i.e., we assume that \Text{\SA{i}..\SA{i}+\max(\LCP{i},\LCP{i + 1})} is stored on one \CN{} for all $1\leq i\leq n$.
This is usually the case if the text \Text{} is composed of a number of smaller documents such that all documents reside on a \emph{single} \CN, but all \CN{}s still have $\Theta\left(\frac{n}{c}\right)$ characters.
(If this is not the case, one could still replicate parts of the text on each \CN{} such that the \CN{}s hold overlapping parts of the text.)
Under this assumption, each \CN{} needs to \emph{send} $\Oh{\frac{n}{c}}$ characters as edge labels. 
Further, each \CN{} also \emph{receives} at most $\Oh{\frac{n}{c}}$ characters, as the local Patricia trie has less than $2\frac{n}{c}$ edges.
Finally, we note that the construction takes one superstep (construct the tree and store the text positions, then retrieve the characters at those positions).
This leads to the following Lemma (where the claim about space follows easily because the local Patricia tries store the edge labels, the \SA{}-values at the leaves, and the skip values at the edges).

\begin{lemma}
  Given that all mismatching characters are stored at the same \CN{} where the corresponding suffix starts, the \SA{}, and the \LCP{}-array, constructing the Patricia tries costs $\Oh{\frac{n}{c}+\frac{n}{c}G+L}$.
  Each $\PT{}_i$ requires $\abs{\text{tree structure}}+\Oh{\frac{n}{c}\left(\lg n +\lg\sigma\right)}$ bits of space.
\end{lemma}

% \subsubsection{Global Trie.}
\paragraph{Global Trie.}
\label{ssub:global_trie}
Next, we consider the construction of the \emph{global trie} \GT{}, which allows us to distribute queries without accessing the text.
\GT{} is the same at every \CN{}, which allows arbitrary \CN{}s to initially process any query.
To identify all \CN{}s that are responsible for a pattern, we require the smallest and largest suffix that is represented by each \CN{} and their \textsf{lcp}-values.
Using this set of suffixes $\mathcal{S}=\lbrace S_{\SAi{1}{1}},S_{\SAi{\frac{n}{c}}{1}},\dots,S_{\SAi{1}{c}},S_{\SAi{\frac{n}{c}}{c}}\rbrace$ to construct \GT{}, we can use the following observation to identify all \CN{}s that are responsible for a pattern.

\begin{observation}\label{obs:query_distribution}
  \CN{} $i$ is responsible for a pattern \Pattern{} if and only if $\Text{\SAi{1}{i}..\SAi{1}{i}+\abs{\Pattern{}}-1}\leq_{\text{lex}} \Pattern{}$ and $\Pattern{}\ \leq_{\text{lex}}\Text{\SAi{\frac{n}{c}}{i}..\SAi{\frac{n}{c}}{i}+\abs{\Pattern{}}-1}$.
\end{observation}
Obviously, there can be pattern for which multiple \CN{}s are responsible.
Depending on the type of query, we need to use the second level index of at most two \CN{}s to answer a query.
Communication with more \CN{}s may be necessary -- see \S~\ref{sub:querying} for more details.

The global trie can be constructed similar to the local Patricia trie construction described above.
The suffixes required for the construction are known (all suffixes in $\mathcal{S}$).
We still require the size of the longest common prefixes of those suffixes.
For two lexicographically consecutive suffixes the size is in the \LCP{}-array.
The size of the longest common prefix of the other suffixes is the string depth of the root of the corresponding local \PT{}.
We can propagate all these values during one communication phase, where each \CN{} sends the two text positions and \textsf{lcp}-values to all other nodes, sending \Oh{c} messages of constant size.
At the end of the phase each \CN{} has a temporary \SA{} and a temporary \LCP{}-array each of size $2c$.
Using these arrays we use the algorithm described above, only handling edge labels differently.

The task of the global trie \GT{} is to distinguish all elements in $\mathcal{S}$ without accessing \Text{}.
Therefore, the edge labels may consist of more than one character.
The first character of an edge $(v,w)$ is the same character we would store if we constructed a Patricia trie.
Let the text position of this character be $i$.
Instead of storing only this character and the string depth, we now need to store the substring \Text{i..i+\strdepth{w}} as the edge label of $(v,w)$.
Hence, it is not necessary to store the string depth at the nodes.  
In addition, we construct the trie with respect to a maximum pattern size $\abs{\Pattern{}}_{\max}$.
We can usually assume that $\abs{\Pattern{}}_{\max}$ is constant (chosen during construction) such that the size $\abs{\Pattern{}}$ of each pattern \Pattern{} is at most $\abs{\Pattern{}}_{\max}$.
Thus, the total size of all edge labels is bounded by $2c\abs{\Pattern{}}_{\max}$.
During the construction, we store references to the edge labels, i.e., the text position and length.
Therefore, at each \CN{} it is known which substrings need to be communicated (as edge labels).
The edge labels are distributed among the \CN{}s in two supersteps.
First, each \CN{} sends an equal amount of different labels to each \CN{}.
The cost for this superstep (including the construction of the tree structure) is $\Oh{c+c\abs{\Pattern{}}_{\max} G+L}$.
In the next superstep, each \CN{} distributes the received labels to each other \CN{}, costing $\Oh{c\abs{\Pattern{}}_{\max} G+L}$.

To prevent that a requested substring spans over more than one \CN{} we pad the locally stored text with the next $\abs{\Pattern{}}_{\max}$ characters.
Since we build the trie for $2c$ substrings, this requires \Oh{c} time, which leads to the following Lemma.

\begin{lemma}
  Given the \SA{} and \LCP{}-array, constructing the global trie costs $\Oh{c+c\abs{\Pattern{}}_{\max}G+L}$.
  The trie requires $\abs{\text{tree structure}}+\Oh{c\abs{\Pattern{}}_{\max}\lg\sigma}$ bits of space.
\end{lemma}

\paragraph{Reducing the Memory Overhead.}
Now we show how we can reduce the memory overhead by increasing the number of supersteps required during construction.
The whole index is kept in main memory, therefore, we want the overhead during the construction to be as small as possible.
First, note that we can stream the \SA{} and \LCP{}-array, since we just need to scan them once for the construction.
Second, we look at the size of the indices, as usually a text position requires more space than a character.
Since we need characters but obtain text positions, we need to store them until the next communication phase.
Usually $\lg n \gg\lg \sigma$ (i.e., factor of five to ten in practice), thus the text positions consume more memory than the labels will later on.
If we only compute $s$ required text positions during a superstep and then retrieve them, we need $\Oh{\frac{n}{sc}}$ supersteps.
Thus, we can decrease the memory overhead by increasing the number of supersteps that are required during the construction.
This yields the following space-time trade-off (regarding the maximum amount of memory required during construction).
\begin{corollary}
  Given the \SA{} and \LCP{}-array, the cost of constructing the Patricia trie is $\Oh{\frac{n}{c}+\frac{n}{c}G+\frac{n}{sc}L}$ if we only allow $s\lg n$ bits additional space.
\end{corollary}
% section space_time (end)
% subsection construction (end)

\subsection{Querying.} % (fold)
\label{sub:querying}
The global trie \GT{} is constructed for the set $\mathcal{S}=\lbrace S_{\SAi{1}{1}},S_{\SAi{\frac{n}{c}}{1}},\dots,S_{\SAi{1}{c}},S_{\SAi{\frac{n}{c}}{c}}\rbrace$ and a maximum pattern length of $\PatternSize_{\max}$.
For any $i\in\lbrack 1,c\rbrack$ the $2i$-th leaf corresponds to the lexicographically smallest suffix and the $2i+1$-th leaf corresponds to the greatest suffix represented by \CN{} $i$.
Querying \GT{} is different from querying a trie as we do not want to find all occurrences of a pattern \Pattern{}, but want to find all \CN{}s that represent \Pattern{}.
We still follow the edges according to their label and the corresponding position in \Pattern{} until we have matched \Pattern{} at a node $u$ or have a mismatch with the label of an edge $\left(v, w\right)$.

In the first case and if $v$ is an internal node, we need to identify the leftmost and rightmost leaves below $v$.
Let $k$ and $\ell$ be the leaf ranks of those leaves, resp.
Then all \CN{}s $j$ with $j\in\lbrack\lfloor\frac{k}{2}\rfloor, \lfloor\frac{\ell}{2}\rfloor\rbrack$ contain positions where the pattern occurs, as the \CN{}s cannot be distinguished by \Pattern{}.
If (in the first case) $v$ is a leaf with leaf rank $k$, then \CN{} ${\lfloor\frac{k}{2}\rfloor}$ can be responsible for \Pattern{}.
In this case we cannot be sure, as \pathlabel{v} may be a prefix of \Pattern{}.
Therefore, we send \Pattern{} to \CN{} ${\lfloor\frac{k}{2}\rfloor}$ and use the local Patricia trie $\PT{}_{\lfloor\frac{k}{2}\rfloor}$ to determine whether \Pattern{} occurs.

In the second case (there was a mismatch), \Pattern{} can still occur.
Let $\alpha$ and $\beta$ be the mismatching characters of the label and the pattern, resp.
If $\alpha>_{\text{lex}}\beta$ we look at the leftmost leaf below $w$.
If the leaf rank $k$ is even, \Pattern{} does not occur in any \CN{}, as it is smaller than the lexicographically smallest suffix represented by \CN{} ${\lfloor\frac{k}{2}\rfloor}$ and greater than the lexicographically greatest suffix represented by \CN{} ${\lfloor\frac{k}{2}\rfloor-1}$ because otherwise another edge would be followed in the beginning.
If the rank is odd, \Pattern{} may occur in \CN{}s ${\lfloor\frac{k}{2}\rfloor}$.
In the other case ($\alpha<_{\text{lex}}\beta$), we need to get the rightmost leaf below $w$ and check the leaf rank.
There may be an occurrence if the leaf rank is even and there cannot be an occurrence if the leaf rank is odd (with the same type of argumentation given before).
All \CN{}s that are responsible for a pattern \Pattern{} form a consecutive interval that we denote by $\GT{\Pattern{}}=\lbrack\ell,r\rbrack$.

\begin{lemma}
  \label{lem:global_trie}
  Given \GT{} and a pattern \Pattern{}.
  Let $\GT{\Pattern{}}=\lbrack\ell,r\rbrack$, if $\ell\neq r$ then \Pattern{} occurs at least once in \CN{}s $\ell$ and $r$ and $\frac{n}{c}$ times in \CN{}s $j$ for all $j\in\left(\ell, r\right)$.
\end{lemma}

Now we take a look at how to answer \emph{pattern matching} queries in the local Patricia tries.
First, we look at the processing of a single query.
Later, we show how the index can be used to answer a batch of queries.

% \subsubsection{Pattern Matching Queries.}
\paragraph{Pattern Matching Queries.}
There are three types of pattern matching queries that the \DPT{} supports:
\begin{description}
\item[Existential:] Given a pattern \Pattern{}, we want to know whether the pattern \Pattern{} occurs in the text \Text{}.
\item[Counting:] Given a pattern \Pattern{}, we want to know how often the pattern \Pattern{} occurs in the text \Text{}.
\item[Enumeration:] Given a pattern \Pattern{}, we want to know all text positions in \Text{} where \Pattern{} occurs.
\end{description}
First, we look at an existential query \Pattern{} that arrives at \CN{} $i$.
We can answer the query in three supersteps (see also Figure~\ref{fig:index_overview}):

\begin{enumerate}
\item At \CN{} $i$, we identify all \CN{}s that are responsible for \Pattern{}, i.e., all \CN{}s $j$ with $j\in\GT{\Pattern{}}$.
If $\ell\neq r$ we know that \Pattern{} occurs in \Text{} (see Lemma~\ref{lem:global_trie}), else we send \Pattern{} to \CN{} $\ell$.
\item Next, we perform a blind search in $\PT{}_\ell$.
If the blind search fails, we know that \Pattern{} does not occur in \Text{}.
Otherwise, the blind search returns a text position $q$.
During the communication phase we retrieve $\Text{q..q+\abs{\Pattern{}}-1}$.
\item Using $\Text{q..q+\abs{\Pattern{}}-1}$ we can verify the existence of \Pattern{} in \Text{} in the third superstep.
When a query can be answered at a \CN{}, we do not send it somewhere else, as the target depends on the application the index is used for.
\end{enumerate}
The cost of an existential query is the following.
During the first superstep we identify all \CN{}s that can answer the query and send it to one \CN{} costing $\Oh{\TimeTrie{\Pattern{}}+\PatternSize G+L}$.
We let \TimeTrie{\Pattern{}} denote the time required to search for \Pattern{} in a trie.
Depending on the implementation this requires \Oh{\PatternSize\lg\sigma} time (binary search) or \Oh{\PatternSize+\lg\lg\sigma} time \cite{Fischer2015a}.
In the second superstep we perform a blind search and retrieve a substring of length \PatternSize.
This costs $\Oh{\TimeTrie{\Pattern{}}+\PatternSize G+L}$.
During the last superstep we just compare two strings of length \PatternSize\ in \Oh{\PatternSize} time.

Counting all occurrences of a pattern can be seen as an extension of the existential query and can be answered similarly (requiring four supersteps).
Let \Pattern{} be a counting query arriving at \CN{} $i$.

\begin{enumerate}
\item First, we identify all \CN{}s that are responsible for \Pattern{}, i.e., all \CN{}s $j$ with $j\in\GT{\Pattern{}}$.
Let all \CN{}s $j$ with $j\in\lbrack\ell,r\rbrack$ be responsible for \Pattern{}.
If $\ell\neq r$ we know that \Pattern{} occurs in all those \CN{}s (see Lemma~\ref{lem:global_trie}).
During the communication phase we send two queries $Q_\ell$ and $Q_r$ to \CN{} $\ell$ and $r$, resp.
The former asks for the lexicographically smallest occurrence of \Pattern{} in $\PT{}_\ell$ and the latter asks for the lexicographically largest occurrence of \Pattern{} in $\PT{}_r$.
\item In the next step, we perform one blind search in $\PT{}_\ell$ and one blind search in $\PT{}_r$.
If $\ell\neq r$ we know that the blind searches will return two text positions $q_\ell$ and $q_r$ that are the lexicographically smallest and largest occurrences of \Pattern{} in \Text{}.
If one of the blind searches fails we know that the \CN{} is not responsible for \Pattern{} and we can send that there are no occurrences at the \CN{}.
During the communication phase, we retrieve $\Text{q_\ell..q_\ell+\abs{\Pattern{}}-1}$ and $\Text{q_r..q_r+\abs{\Pattern{}}-1}$.
\item Using $\Text{q_\ell..q_\ell+\abs{\Pattern{}}-1}$ and $\Text{q_r..q_r+\abs{\Pattern{}}-1}$ we can verify the existence of \Pattern{} in \Text{} (only necessary if $\ell\neq r$) and also find the number of occurrences at \CN{}s $\ell$ and $r$ using the leaf ranks.
We send the number to \CN{} $i$.
\item We know the number of occurrences $occ_\ell$ and $occ_r$ of \Pattern{} in \CN{}s $\ell$ and $r$, resp.
We also know that \Pattern{} has to occur $\frac{n}{c}$ times at each \CN{} $j$ for $j\in\left(\ell, r\right)$.
Thus the total number of occurrences of \Pattern{} is $occ_\ell+occ_r+\max\left(0, r-\ell-1\right)\frac{n}{c}$.
\end{enumerate}
The first superstep costs \Oh{\TimeTrie{\Pattern{}}+\PatternSize G+L} as we need to identify the \CN{}s that are responsible for the pattern and send it to two \CN{}s.
In the second superstep we perform a blind search at two \CN{}s and retrieve two substrings of length \PatternSize.
This costs $\Oh{\TimeTrie{\Pattern{}}+\PatternSize G+L}$.
The third superstep consist of comparing the retrieved substrings with the pattern and send the number of occurrences of the pattern to the \CN{} where the pattern arrived initially, i.e., \CN{} $i$.
This costs \Oh{\PatternSize + G + L}.
During the last superstep we need to compute the total number of occurrences at \CN{} $i$ which costs \Oh{1+L}.

Last, we consider enumeration queries.
Let \Pattern{} be a enumeration query arriving at \CN{} $i$.
During the first two supersteps answering an enumeration query does not differ from answering a counting query.
The remaining steps are changed as follows.
\begin{enumerate}
\setcounter{enumi}{2}
\item Using $\Text{q_\ell..q_\ell+\abs{\Pattern{}}-1}$ and $\Text{q_r..q_r+\abs{\Pattern{}}-1}$ we can verify the existence of \Pattern{} in \Text{} (only necessary if $\ell\neq r$) and also find all positions where the pattern occurs.
We send all these positions to \CN{} $i$.
\item We have received all occurrences of \Pattern{} from the \CN{}s $\ell$ and $r$.
Next we need to retrieve all occurrences, i.e., the local \SA{} from all \CN{}s $j$ for $j\in\left(\ell,r\right)$.
\end{enumerate}
The first two supersteps are the same as for a counting query.
Therefore, the costs of the first two supersteps are the same.
The third superstep is very similar to the third superstep for answering a counting query.
The only difference is that we need to send the text positions of all occurrences to the \CN{} where the query arrived initially.
This costs \Oh{\PatternSize + occ\cdot G + L}, where $occ$ denotes the number of occurrences of \Pattern{} at \CN{} $\ell$ and \CN{} $r$.
Last, we need to retrieve the text positions of all occurrences of \Pattern{} in \CN{}s $j$ with $j\in (\ell,r)$, which costs \Oh{(r-\ell)\frac{n}{c} G + L} if $\ell< r+1$.

Answering any type (existential, counting or enumeration) of query has (asymptotically) the same cost for the first two supersteps, as we send at most twice as many queries to the second level (for counting and enumeration queries).
To answer counting queries we send the leaf ranks during third superstep yielding a cost of \Oh{\PatternSize + G+L}.
In the last superstep we just need to add up the number of occurrences in \Oh{1} time.
When we consider enumeration queries, we need to report all text positions where \Pattern{} occurs.
Let $occ$ be the maximum number of occurrences of \Pattern{} in a \CN{} $j$ for $j\in\lbrack\ell,r\rbrack$, then the cost of the third superstep is \Oh{\PatternSize+occ\cdot G+L}.
In the fourth superstep we need to retrieve all positions from the \CN{}s $j$ for all $j\in\left(\ell,r\right)$ costing \Oh{1+c\cdot occ\cdot G+L}.
All in all we get the following costs.

\begin{lemma}
  Given a pattern \Pattern{} answering an existential query costs
  \Oh{\TimeTrie{\Pattern{}}+\PatternSize G+L}, answering a counting
  query costs \Oh{\TimeTrie{\Pattern{}}+ \PatternSize G+L}, and answering an enumeration query costs
  \Oh{\TimeTrie{\Pattern{}}+\left(\PatternSize+occ\right) G+L}, where $occ$ denotes the total number of
  occurrences of \Pattern{}.
\end{lemma}

% \subsubsection{Batched Queries and Load Balancing} % (fold)
\paragraph{Batched Queries and Load Balancing.}
\label{ssub:batched_queries}
When we process a batch of $q$ queries at once rather than a single query, the number of supersteps does not increase, i.e., we can amortize the startup latencies of the BSP model over a large number of queries.
Moreover, if the local work and communication volume is well balanced over the \CN{}s, the query throughput scales linearly with $c$.
On the one hand, we can measure empirically how well balanced the computation is -- see also \S~\ref{sec:experiments}.
On the other hand, we can see to what extent good balance can be enforced.

Balancing the queries itself can be achieved using any standard load balancing technique, i.e., assuming that $\Oh{q/c}$ queries arrive at each \CN{} is unproblematic.

Balancing how many queries get directed at each local trie is more difficult, since certain patterns might be more popular than others.
However, we can use ``virtualization'' -- we split the corpus into $c'\gg c$ pieces and distribute them randomly to the actual \CN{}.

Similarly, some documents might be more popular than others.
However, by randomly permuting the documents in the corpus, we can at least ensure that it is unlikely that many popular documents are assigned to the same \CN{}.

When all these balancing conditions are fulfilled, a batch $Q$ of queries can be completed in time $$\Oh{\frac{1}{c}\left(\sum_{\Pattern{}\in Q}\TimeTrie{\Pattern{}}+(\abs{\Pattern{}}+occ(\Pattern{}))G\right)+L},$$ where $occ(\Pattern{})$ denotes the number of occurrences of $\Pattern{}$ for an enumeration query (and $0$ else).

% \subsubsection{Comparison to the Distributed Suffix Array}
\paragraph{Comparison to the Distributed Suffix Array.}
Using the (multiplexed) distributed suffix array (\DSA{})~\cite{Arroyuelo2014}, a batch $Q$ of $q$ counting queries can be answered in time $$\Oh{\frac{1}{c}\left(\sum_{\Pattern{}\in Q}t_{\text{Bin}}(\Pattern{})+(\abs{\Pattern{}}\lg\frac{n}{q}+\lg c)G\right)+\lg\frac{cn}{q}L},$$ where $t_{\text{Bin}}(\Pattern{})$ denotes the time to identify the occurrences of the pattern \Pattern{} in the \SA{}, i.e., $\abs{\Pattern{}}\lg n$.
Distributing the queries costs \Oh{\sum_{\Pattern{}\in Q}\abs{\Pattern{}}+\abs{\Pattern{}}G+L} and is dominated by the costs of answering the batch of queries.

Comparing the costs of the \DSA{} with our \DPT{} we get the following result:
The maximum time used for computation by each \CN{} is \Oh{\frac{1}{c}\left(\sum_{\Pattern{}\in Q}\TimeTrie{\Pattern{}}\right)} using the \DPT{} since we look in \GT{} and at most two local \PT{}s for each query.
The computation time required by the \DSA{} is \Oh{\frac{1}{c}\left(\sum_{\Pattern{}\in Q}t_{\text{Bin}}(\Pattern{})\right)} and results from the binary searches (local and inter-\CN{}).
Hence, the time used for computation by each \CN{} differs with respect to the time required for searching the corresponding suffix array interval for each pattern using a trie and using binary search.  
Usually, we can assume that \TimeTrie{\Pattern{}} is smaller than $t_{\text{Bin}}(\Pattern{})$.

The cost of communication is \Oh{\frac{1}{c}\left(\sum_{\Pattern{}\in Q}(\abs{\Pattern{}})\right)G} using the \DPT{}, as we just send each pattern to at most two \CN{}s and retrieve a substring of the length of the pattern.
For the \DSA{} the cost of communication is higher, i.e., \Oh{\frac{1}{c}\left(\sum_{\Pattern{}\in Q}\abs{\Pattern{}}\lg\frac{n}{q}+\lg c\right)G} because more substrings need to be retrieved during the binary search.
This effect can be moderated by storing pruned suffixes for each position of \SA{} -- see \S~\ref{sub:pruning_dsa}.
Still, the \DPT{} requires only a constant number of substrings to be retrieved for each query.

Last, the synchronization using \DPT{} is constant, i.e., \Oh{L}, but using the \DSA{} synchronization costs \Oh{\lg\frac{cn}{q}L}.
Due to the constant number of messages being sent using the \DPT{}, the synchronization cost is optimal and a logarithmic factor worse using the \DSA{}.

Therefore, if we assume an optimal distribution of the queries and of the documents, the \DPT{} is theoretically faster than the \DSA{}.
This difference in cost can also be seen in practice -- see \S~\ref{sec:experiments}.
However, the multiplexed \DSA{} is very strong against query bias, whereas the \DPT{} can be affected by query bias resulting in a load imbalance.
% subsubsection batched_queries (end) subsection querying (end)
% queries_in_the_distributed_patricia_trie (end)

\section{Experiments} % (fold)
\label{sec:experiments}
We implemented the distributed Patricia trie using C{}\verb!++! (\texttt{g}{}\verb!++! 6.1 with flags \texttt{-03} \texttt{-march=native}).
The communication is handled by the \emph{Message Passing Interface} (MPI, Open MPI 1.10.3) with each MPI process representing a \CN{} of our algorithm.
For the representation of bit vectors and the operations \rank{}{}, \select{}{} and \findclose{\,\!}, we use the \emph{succinct data structure library} (sdsl-lite, 2.0.1) \cite{Gog2014}.
In particular we used the \textsc{rank\_support\_v5} for \rank{}{}, \textsc{select\_support\_mcl} for \select{}{} and \textsc{bp\_support\_sada} for \findclose{\,\!}.
We computed the \SA{} and \LCP{}-array using the implementation of Flick et al.~\cite{Flick2015}.
The source code of our implementation and data required to reproduce our results are available from \url{https://github.com/kurpicz/dpt}.

For our experiments we use the \emph{common crawl corpus} as input.\footnote{\url{http://commoncrawl.org/}} It provides world wide web crawl data and contains raw content, text only and metadata of the crawled websites from about 1.23 billion web pages.
In total the corpus has a size of 541 TB (as of 27.07.2016).
We use the \emph{WET files} that contain only the text without any tags or other meta information.
(We removed all additional data added by the common crawl corpus.)

The data we use for queries comes from the following sources:
\begin{description}
\item[Text Retrieval Conference (TREC)] Containing all published test queries of the \emph{Million Query Track}, which contains 60k queries in total~\cite{MillionQueryTrack}.
\item[AOL Query Log (AOL)] Contains around 20M web queries collected from roughly 650k users that have been collected over three months \cite{Pass2006}.
\end{description}

To the best of our knowledge, the distributed suffix array (\DSA{}) presented by Arroyuelo et al.~\cite{Arroyuelo2014} is the only other (available) implementation of a distributed full-text index.
Hence, we compare our results with the fastest variant of the \DSA{}, the \emph{multiplexed} distribution of the \SA{}.
The implementation of the \DSA{} only supports counting queries. Therefore, in \S~\ref{sub:construction_and_query_time} we only compare the time for counting queries.
We adapted the (non-public) source code such that it works with texts of size greater than 4~GiB.

All experiments were conducted on the \emph{InstitutsCluster II} (IC2) at KIT.\footnote{\url{https://www.scc.kit.edu/dienste/ic2.php}}
The cluster has compute nodes  consisting of two Octa-Core Intel Xeon E5-2670 processors with 64~GB main memory and 2~TB external memory.
The nodes are connected using InfiniBand QDR.
Up to 32 nodes at a time were available for our measurements.
Each node runs 16 MPI processes and in this section \CN{} refers to MPI process.

\subsection{Construction.}
\label{sub:experiments_construction}
\begin{figure}
  \centering
  \includegraphics[scale=1]{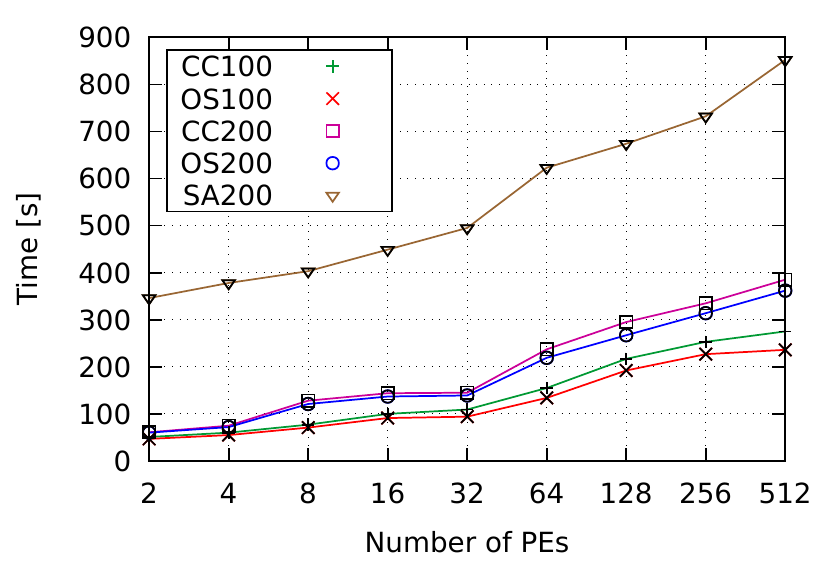}
  \caption{Construction times of the \DPT{} for 100~MiB (\textsf{CC100} and \textsf{OS100}) and 200~MiB (\textsf{CC200} and \textsf{OS200}) of text per \CN{}, utilizing collective communication (\textsf{CC}) and one-sided communication (\textsf{OS}).
  \textsf{SA200} is the construction time of the \SA{} and \LCP{}-array for 200~MiB text per \CN{}.
  \label{fig:construction_time_weak}}
\end{figure}

As noted before, the construction of the \SA{} and \LCP{}-array are not part of this paper -- we start the timer as soon as each \CN{} holds its local share of the \SA{}, \LCP{}-array and \Text{}.
Still, the construction of the \SA{} and \LCP{}-array are the bottleneck of our data structure.
Figure~\ref{fig:construction_time_weak} shows the construction time of the distributed Patricia trie using a \emph{weak scaling} plot, i.e., for each curve the amount of input data per \CN{} is fixed.
In this case we used up to 100~GiB of text which translates to 200~MiB of text per \CN{} and constructed the \DPT{} for $\abs{\Pattern{}}_{\max} = 30$.
We could not run bigger experiments as we were only able to compute the \SA{} and \LCP{}-array for texts up to that size using the available resources and the only distributed construction algorithm that can compute both arrays, i.e.,~\cite{Flick2015}.

The construction time for the \SA{} and \LCP{}-array is two to three times greater than the time required for the construction of the \DPT{}.
Hence, we can say that our construction time is reasonable and practical.

We compare two different variants of implementation: Using the collective communication operation \textsc{MPI\_Alltoallv} and using a large batch of \emph{remote direct memory access} operations (RDMA, operation \textsc{MPI\_Get}).
Preparing the requests and constructing the tree topology are the most time consuming tasks during the construction of the \DPT{}.
One-sided communication turns out to be somewhat faster since it does not need to send text positions first, but can directly access the text.
The amount of messages sent and received by all \CN{}s is (with a few small exceptions) homogeneously distributed among all \CN{}s -- see Figure~\ref{fig:distribution_construction_queries} (left).

The space consumption of our index can be seen in Table~\ref{tab:space_pt}, where we compare the different representations of the local Patricia tries.
Since we need to keep track of the nodes on the stack, the memory usage is higher during the construction.
In the following, we omit a detailed analysis of DFUDS and BP, as LOUDS is better with respect to size and speed.

The multiplexed distributed suffix array by Arroyuelo et al.~\cite{Arroyuelo2014} requires preprocessing of \SA{} and \Text{}.
During the preprocessing, the multiplexed \SA{} for each \CN{} is constructed and pruned suffixes for each text positions are stored accordingly (see \S~\ref{sub:pruning_dsa} for the effect of pruning in practice).
We refer to this preprocessing as construction time.
However, the process is not distributed; one \CN{} prepares all required files.
Therefore, we omit a comparison of the construction time with \DPT{} as it does not scale.

A standard question for a parallel algorithm is about its speedup with respect to the best sequential algorithm.
This is important for understanding how much overhead is involved in going to a distributed environment.
Since the \SA{} and \LCP{}-array construction dominates index construction time and since there is no tuned 
sequential implementation of \DPT{} construction itself, we make this comparison only for the \SA{} and \LCP{}-array construction.
The fastest sequential algorithm we are aware of is \emph{divsufsort}, which computes the \SA{} (but not the LCP array).\footnote{\url{https://github.com/y-256/libdivsufsort}}
Running the 64-bit version of divsufsort 2.0.1 for 50~GiB on a machine with 512~GB RAM and 4 Intel Xeon E5-4640 processors takes 42\,247 seconds.
Extrapolating this to 100~GiB gives 84\,494 seconds -- about 100 times more than the algorithm from~\cite{Flick2015}. 
Note that divsufsort is much slower when extrapolating from traditionally small inputs as we need a 64-bit version, due to  NUMA effects, and because for really large inputs a logarithmic term in virtual address translation becomes noticeable~\cite{jurkiewicz2015model}.
There exists also a parallel variant of divsufsort, where one of its two sorting steps is parallelized.
Hence, not the whole algorithm is executed in parallel.
Therefore, on 50~GiB and using 32 cores, it is only about 25\% faster than sequential divsufsort.

Another interesting comparison is with a state of the art general purpose external memory algorithm \cite{BFO16}. Here, the algorithm from~\cite{Flick2015} on 512 processors is about 280 times (using somewhat faster processors but on larger inputs than~\cite{BFO16}).

\begin{table}
  \centering
  \begin{tabular}{lllll}
    \toprule
       & Pointer & BP & DFUDS & LOUDS \\
    \midrule
    peak & $46n$ & $19n$ & $18n$ & $18n$\\
    \midrule
    size & $42n$ & $16n$ & $16n$ & $15n$\\
    \bottomrule
  \end{tabular}
  \caption{Bits needed to store the local tries during construction (peak) and querying (size) using 40-bit text positions.\label{tab:space_pt}}
\end{table}
% subsection distribution_of_the_communication (end)

\subsection{Query Time.}
\label{sub:construction_and_query_time}
In this section we focus on pattern matching queries.
Once more, we perform a weak scaling experiment to compare the times required to answer batches of queries.
This time, both the input size per \CN{} (200~MiB) and the number $q$ of queries arriving at each \CN{} (20k, 40k and 80k) are fixed.
Thus, we build an index on up to 100~GiB of text, and want to answer a batch of at most 41M queries.
If the number of queries exceeds the available number of queries (60k and 20M for TREC and AOL queries, resp.), we replicate the set of queries accordingly.
Also, we stop the timing as soon as the result for a query is known at any \CN{}.
For example, if we have a counting query that can be answered by a single \CN{}, we do not send the result to the \CN{} where the query arrived initially.

\begin{figure*}[htb]
  \centering
  \includegraphics[scale=.9125]{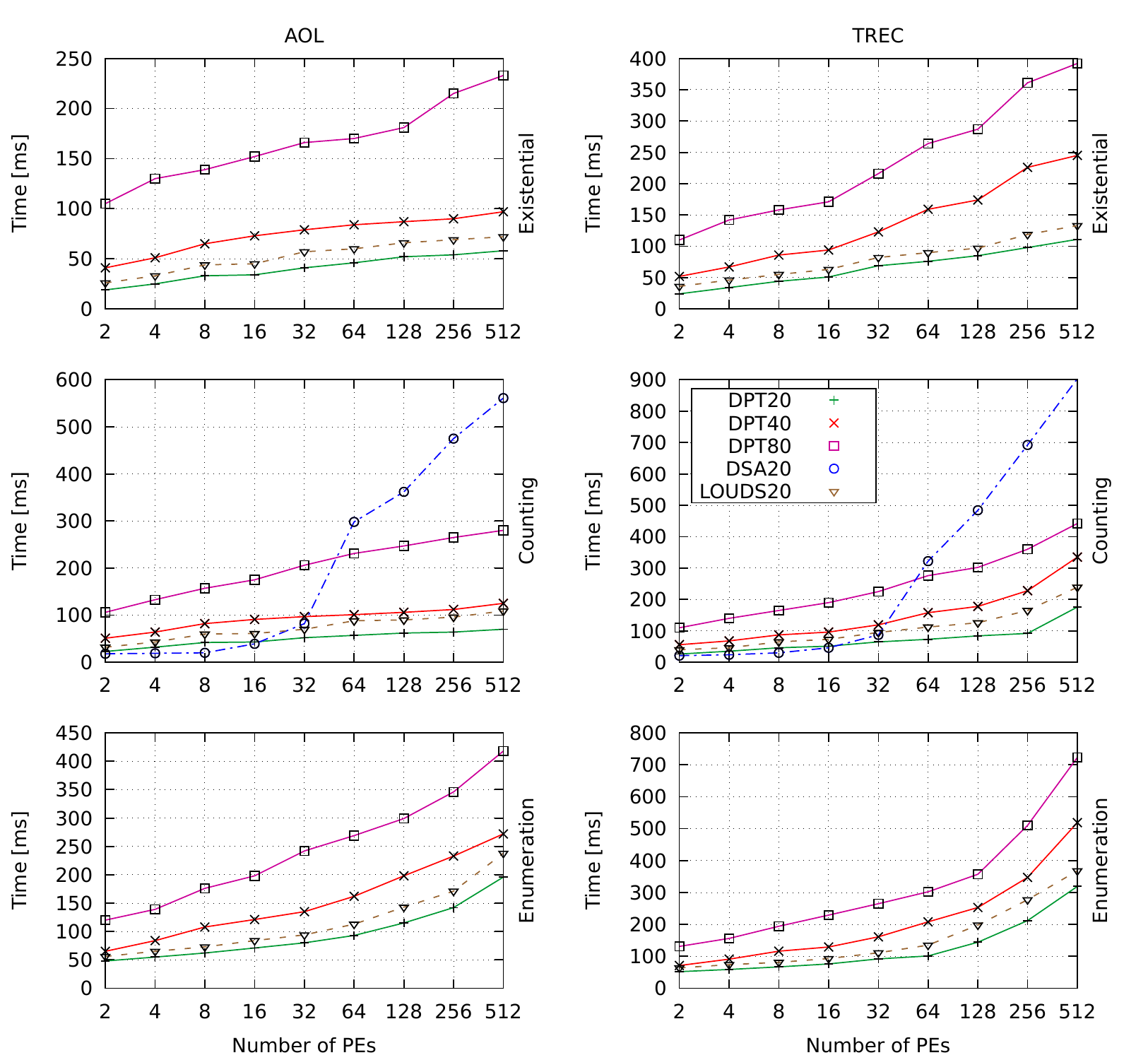}
  \caption{Query times of the \DPT{} for 200~MiB of text per \CN{}.
  Each \CN{} receives 20k (\textsf{DPT20}), 40k (\textsf{DPT40}) and 80k (\textsf{DPT80}) queries.
  \textsf{LOUDS20} is the query time for 20k queries using LOUDS as succinct trie representation and \textsf{DSA20} is the query time for the distributed suffix array, also asking for 20k queries per \CN{}.\label{fig:query_time}}
\end{figure*}

Figure~\ref{fig:query_time} shows the query times for existential, counting and enumeration queries.
\textsf{DPT{}20} represents the times required to answer a batch of 20k queries per \CN{}, \textsf{\DPT{}40} and \textsf{\DPT{}80} denote the times for 40k and 80k queries, using the pointer-based representation of the tries.
The labels \textsf{\DSA{}20} and \textsf{LOUDS20} denote the times for 20k queries using the \DSA{} and the succinct LOUDS representations of the \DPT{}.
We augment the \DSA{} with pruned suffixes of size 5 -- see \S~\ref{sub:pruning_dsa} for more details.
All query types scale reasonably well with the number of \CN{}s.
Existential queries can be answered the fastest, as they only require a single blind search (including the retrieval of a substring).
They can be answered up to twice as fast as the counting queries.

Counting queries require more work at each \CN{}, as we need to compute the number of occurrences, which translates to two traversals from one node to a leaf at the local \PT{}.
Comparing our implementation with the \DSA{} we see that we scale better as we can send our queries to the \CN{} that can actually answer them.
This allows us to answer queries up to 5.5 times faster.
The implementation of \DSA{} only supports counting queries, therefore, \DSA{} only appears for counting queries in Figure~\ref{fig:query_time}.
When utilizing more than 32 \CN{}s, \DSA{} becomes significantly slower than \DPT{}.
This is probably due to the increasing number of messages sent by the index that are necessary during the binary search, whenever the pattern must be compared with the text.
For larger texts, there are fewer text positions that correspond to text that is locally available.
This effect can be reduced by \emph{pruning} (see \S~\ref{sub:pruning_dsa}).

Last, we have enumeration queries, which are the hardest to answer as we have to compute all text positions.
They can take up to 1.5 times as long as counting and 3 times as long as existential queries.

\begin{figure*}
  \centering
  \includegraphics[scale=1]{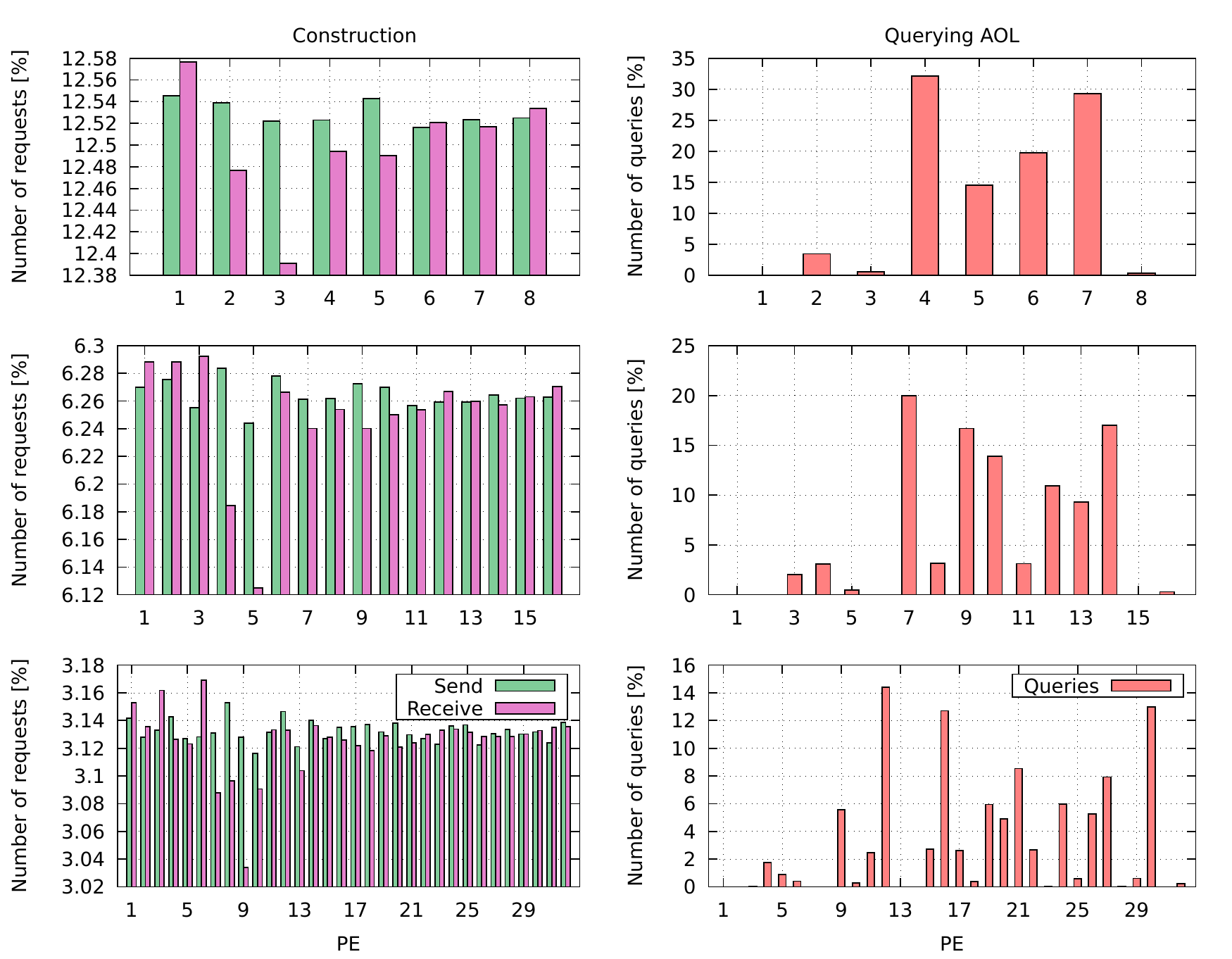}
  \caption{Work imbalance between different PEs. For each PE, we give the percentage of requests sent and received during construction of the \DPT{} and percentage of queries received during querying by each \CN{}.
    Here, the \DPT{} is constructed for 200~MiB of text per \CN{} and we ask for 20k AOL queries per \CN{}.}
  \label{fig:distribution_construction_queries}
\end{figure*}

The TREC queries are generally harder to answer as there are fewer and we have duplicates as soon as we require more than 60k queries.
Duplicates lead to a higher imbalance regarding the query distribution, i.e., query bias -- see Figure~\ref{fig:distribution_construction_queries} (right) for the query distribution among the \CN{}s.

The succinct tree representations are reasonably fast compared with the pointer based implementation -- see again Figure~\ref{fig:query_time}.
LOUDS is the fastest of the three tested succinct tree representations, being only 10\% to 20\% slower than the pointer based implementation but using only about a third of the space.
This means that succinct data structures allow us to use 2.5 times less memory to build the \DPT{} for the same text as a pointer based implementation. 
Counting and enumeration queries are more expensive when it comes to succinct tree structures, as we need to traverse the trie multiple times if the pattern occurs.
% subsection construction_and_query_time (end)

\subsection{Distributed Suffix Array -- Pruning.}
\label{sub:pruning_dsa}
In addition to the local multiplexed \SA{}, the (multiplexed) \DSA{} also holds \emph{pruned suffixes}, i.e., for each position in the multiplexed \SA{}, the first $\ell$ characters of the corresponding suffix are stored.
Since the text corresponding to a suffix may not be available locally, this speeds up the query time (by increasing the size of the index).
We tested different sizes $\ell$ for the pruned suffixes, using 200~MiB of text per \CN{} and asking for 20k AOL queries per \CN{}, i.e., the same configuration as in our experiments for the query times.
The speedup is listed in Table~\ref{tab:speedup_pruned_dsa}.
\begin{table}[ht]
  \centerline{%
    \begin{tabular}{llllll}
      \toprule
      \CN{}s & $\ell$ & speedup & \CN{}s & $\ell$ & speedup\\
      \midrule
      2 & 0 & 1 & 8 & 0 & 1\\
      \cmidrule{2-3}\cmidrule{5-6}
             & 5 & 1.04 && 5 &1.13\\
      \cmidrule{2-3}\cmidrule{5-6}
             & 10 & 1.20 && 10 &1.51\\
      \cmidrule{2-3}\cmidrule{5-6}
             & 15 & 1.28 && 15 &1.54\\
      \cmidrule{2-3}\cmidrule{5-6}
             & 20 & 1.33 && 20 &1.71\\
      \midrule
      4 & 0 & 1 & 16  & 0 &1\\
      \cmidrule{2-3}\cmidrule{5-6}
             & 5 & 1.33 && 5 &1.05\\
      \cmidrule{2-3}\cmidrule{5-6}
             & 10 & 1.42 && 10 &1.67\\
      \cmidrule{2-3}\cmidrule{5-6}
             & 15 & 1.41 && 15 &1.75\\
      \cmidrule{2-3}\cmidrule{5-6}
             & 20 & 1.45 && 20 &1.92\\
      \bottomrule
    \end{tabular}
  }
  \caption{Speedup of the query time in the \DSA{} with respect to the size $\ell$ of the pruned suffixes.\label{tab:speedup_pruned_dsa}}
\end{table}

Since the average size of an AOL query is 18, pruned suffixes of size greater than 20 do not provide any more significant speedup.
In our experiments we used pruned suffixes of size 5, as this adds 40 bytes to the \DSA{} per text position, which corresponds to the size of our pointer based \DPT{} representation.
Choosing larger pruned suffixes can speedup the \DSA{} such that it is faster than the \DPT{} on 32 \CN{}s.
However, this also results in an index that is larger than our \DPT{}.
%section experiments (end)

\section{Conclusion and Future Work}
\label{sec:conclusion_and_future_work}
We presented a distributed full-text index that supports existential, counting and enumeration queries.
All queries can be answered using a constant number of messages with length proportional to the queries.
Our implementation scales well regarding the query processing and is faster than the \DSA{} when run on more than 32 \CN{}s.
For 512 \CN{}s our index can answer a batch of counting queries up to 5.5 times faster than the \DSA{}.
Also, we use succinct data structures, resulting in lower space requirements by a factor of 2.5 at only about 20\% slowdown.

Still, there are optimizations that may lead to better performance in the future.
One important issue is to further develop and implement the load balancing measures outlined in \S~\ref{ssub:batched_queries}.
An orthogonal issue is to consider a stream of queries, i.e., instead of a batch of queries arriving at each \CN{} at the same time, we assume that queries can arrive at any \CN{} at any time.
This is a real world problem, as for many  applications, queries do not arrive in a batch.
For this setting, we need  asynchronous communication between the \CN{}s.
In this scenario, an interesting question is at which amount of queries batched query processing becomes more efficient than the streaming mode.
%Also, we want to make the \DPT{} more robust against query bias by using techniques such as ``virtualization.''

In addition, our index can be extended to also answer document retrieval queries \cite{navarro14spaces}, where the text is composed of a number of (short) documents, and one wishes to count or enumerate all \emph{documents} containing a given query pattern (documents containing the pattern multiple times should only be counted/enumerated \emph{once}). For document counting, one could use the preprocessing by Hui \cite{Hui1992}, while for optimal document listing, the technique of Hon et al.\ \cite{Hon2014} could be adapted. This latter technique relies on a data structure for fast range minimum queries \cite{fischer11space}, which has to be transformed into a distributed environment.

Another interesting direction is the usage of compressed cache-oblivious tries as shown by Ferragina and Venturini~\cite{Ferragina2013}.
Further improvements should also be possible using hybrid parallelism, i.e., to exploit that \CN{}s on the same compute node can quickly interact using shared memory.

\section{Acknowledgments}
We would like to thank the authors of~\cite{Arroyuelo2014} for providing their source code, and in particular Veronica Gil-Costa for answering all questions regarding the implementation.
We also want to thank Timo Bingmann for providing numbers on sequential suffix array construction.
Our research was supported by the German Research Foundation (DFG), priority programme ``Algorithms for Big Data'' (SPP 1736).


\begin{thebibliography}{99}

\bibitem{Arroyuelo2014}
Diego Arroyuelo, Carolina Bonacic, Veronica Gil-Costa, Mauricio Marin, and
  Gonzalo Navarro.
\newblock Distributed text search using suffix arrays.
\newblock {\em Parallel Computing}, 40(9):471--495, 2014.

\bibitem{Benoit2005}
David Benoit, Erik~D. Demaine, J.~Ian Munro, Rajeev Raman, Venkatesh Raman, and
  S~Srinivasa Rao.
\newblock Representing trees of higher degree.
\newblock {\em Algorithmica}, 43(4):275--292, 2005.

\bibitem{BFO16}
Timo Bingmann, Johannes Fischer, and Vitaly Osipov.
\newblock Inducing suffix and lcp arrays in external memory.
\newblock {\em J. Exp. Algorithmics}, 21:2.3:1--2.3:27, 2016.

\bibitem{Clark1996}
David Clark.
\newblock {\em Compact Pat Trees}.
\newblock PhD thesis, University of Waterloo, 1996.

\bibitem{Clifford2005}
Rapha{\"{e}}l Clifford.
\newblock Distributed suffix trees.
\newblock {\em J. Discrete Algorithms}, 3(2-4):176--197, 2005.

\bibitem{Ferragina1999}
Paolo Ferragina and Roberto Grossi.
\newblock The string b-tree: a new data structure for string search in external
  memory and its applications.
\newblock {\em J. ACM}, 46(2):236--280, 1999.

\bibitem{Ferragina1999a}
Paolo Ferragina and Fabrizio Luccio.
\newblock String search in coarse-grained parallel computers.
\newblock {\em Algorithmica}, 24(3-4):177--194, 1999.

\bibitem{Ferragina2013}
Paolo Ferragina and Rossano Venturini.
\newblock Compressed cache-oblivious string b-tree.
\newblock In {\em Annual European Symposia on Algorithms (ESA)}, volume 8125 of
  {\em LNCS}, pages 469--480. Springer, 2013.

\bibitem{Fischer2015a}
Johannes Fischer and Pawel Gawrychowski.
\newblock Alphabet-dependent string searching with wexponential search trees.
\newblock In {\em Annual Symposium on Combinatorial Pattern Matching (CPM)},
  volume 9133 of {\em LNCS}, pages 160--171, 2015.

\bibitem{fischer11space}
Johannes Fischer and Volker Heun.
\newblock Space efficient preprocessing schemes for range minimum queries on
  static arrays.
\newblock {\em SIAM J.\ Comput.}, 40(2):465--492, 2011.

\bibitem{Flick2015}
Patrick Flick and Srinivas Aluru.
\newblock Parallel distributed memory construction of suffix and longest common
  prefix arrays.
\newblock In {\em International Conference for High Performance Computing,
  Networking, Storage and Analysis (SC)}, pages 16:1--16:10. ACM, 2015.

\bibitem{Gog2014}
Simon Gog, Timo Beller, Alistair Moffat, and Matthias Petri.
\newblock From theory to practice: Plug and play with succinct data structures.
\newblock In {\em International Symposium on Experimental Algorithms (SEA)},
  volume 8504 of {\em LNCS}, pages 326--337. Springer, 2014.

\bibitem{Hon2014}
Wing{-}Kai Hon, Rahul Shah, Sharma~V. Thankachan, and Jeffrey~Scott Vitter.
\newblock Space-efficient frameworks for top-\emph{k} string retrieval.
\newblock {\em J. {ACM}}, 61(2):9:1--9:36, 2014.

\bibitem{Hui1992}
Lucas Chi~Kwong Hui.
\newblock Color set size problem with application to string matching.
\newblock In {\em Annual Symposium on Combinatorial Pattern Matching (CPM)},
  volume 644 of {\em LNCS}, pages 230--243. Springer, 1992.

\bibitem{Jacobson1989}
Guy Jacobson.
\newblock Space-efficient static trees and graphs.
\newblock In {\em Annual Symposium on Foundations of Computer Science (FOCS)},
  pages 549--554. {IEEE} Computer Society, 1989.

\bibitem{jurkiewicz2015model}
Tomasz Jurkiewicz and Kurt Mehlhorn.
\newblock On a model of virtual address translation.
\newblock {\em J. of Experimental Algorithmics}, 19:1.9:1--1.9:28, 2015.

\bibitem{Karkkainen2006}
Juha K{\"{a}}rkk{\"{a}}inen, Peter Sanders, and Stefan Burkhardt.
\newblock Linear work suffix array construction.
\newblock {\em J. {ACM}}, 53(6):1--19, 2006.

\bibitem{Kulla2007}
Fabian Kulla and Peter Sanders.
\newblock Scalable parallel suffix array construction.
\newblock {\em Parallel Computing}, 33:605--612, 2007.
\newblock Special issue on Euro PVM/MPI 2006, distinguished paper.

\bibitem{Larsson2007}
N.~Jesper Larsson and Kunihiko Sadakane.
\newblock Faster suffix sorting.
\newblock {\em Theor. Comput. Sci.}, 387(3):258--272, 2007.

\bibitem{Makinen2015}
Veli M{\"{a}}kinen, Djamal Belazzougui, Fabio Cunial, and Alexandru~I. Tomescu.
\newblock {\em Genome-Scale Algorithm Design: Biological Sequence Analysis in
  the Era of High-Throughput Sequencing}.
\newblock Cambridge University Press, 2015.

\bibitem{Makinen2004}
Veli M{\"{a}}kinen, Gonzalo Navarro, and Kunihiko Sadakane.
\newblock Advantages of backward searching -- efficient secondary memory and
  distributed implementation of compressed suffix arrays.
\newblock In {\em International Symposium on Algorithms and Computation
  (ISAAC)}, volume 3341 of {\em LNCS}, pages 681--692. Springer, 2004.

\bibitem{Manber1993}
Udi Manber and Eugene~W. Myers.
\newblock Suffix arrays: {A} new method for on-line string searches.
\newblock {\em {SIAM} J. Comput.}, 22(5):935--948, 1993.

\bibitem{Morrison1968}
Donald~R. Morrison.
\newblock {PATRICIA} - practical algorithm to retrieve information coded in
  alphanumeric.
\newblock {\em J. {ACM}}, 15(4):514--534, 1968.

\bibitem{Munro2001}
J.~Ian Munro and Venkatesh Raman.
\newblock Succinct representation of balanced parentheses and static trees.
\newblock {\em {SIAM} J. Comput.}, 31(3):762--776, 2001.

\bibitem{navarro14spaces}
Gonzalo Navarro.
\newblock Spaces, trees, and colors: The algorithmic landscape of document
  retrieval on sequences.
\newblock {\em ACM Comput.\ Surv.}, 46(4):Article No.~52, 2014.

\bibitem{MillionQueryTrack}
National~Institute of~Standards and Technology.
\newblock Million query track, 2010.
\newblock \url{http://trec.nist.gov/data/million.query.html}~accessed
  04.07.2016.

\bibitem{Pass2006}
Greg Pass, Abdur Chowdhury, and Cayley Torgeson.
\newblock A picture of search.
\newblock In {\em International Conference on Scalable Information Systems
  (INFOSCALE)}. ACM, 2006.

\bibitem{Russo2010}
Lu{\'{\i}}s M.~S. Russo, Gonzalo Navarro, and Arlindo~L. Oliveira.
\newblock Parallel and distributed compressed indexes.
\newblock In {\em Symp. on Combinatorial Pattern Matching (CPM)}, volume 6129
  of {\em LNCS}, pages 348--360. Springer, 2010.

\bibitem{Sadakane2010}
Kunihiko Sadakane and Gonzalo Navarro.
\newblock Fully-functional succinct trees.
\newblock In {\em Annual {ACM-SIAM} Symposium on Discrete Algorithms, (SODA)},
  pages 134--149. {SIAM}, 2010.

\bibitem{Skillicorn1997}
David~B. Skillicorn, Jonathan M.~D. Hill, and William~F. McColl.
\newblock Questions and answers about {BSP}.
\newblock {\em Scientific Programming}, 6(3):249--274, 1997.

\bibitem{Valiant1990}
Leslie~G. Valiant.
\newblock A bridging model for parallel computation.
\newblock {\em Commun. {ACM}}, 33(8):103--111, 1990.

\end{thebibliography}
\end{document}